\documentclass[a4paper,11pt]{article}

\usepackage[left=2.5cm,right=2.5cm,top=2.5cm,bottom=2.5cm]{geometry}
\usepackage{graphicx,amssymb,amsmath,amsthm,mathrsfs,setspace,subcaption,cite,authblk}
\usepackage[all]{xy}

\usepackage[colorlinks=true,bookmarks=true]{hyperref}
\usepackage{breakurl} 

\usepackage{mathtools}

\DeclarePairedDelimiterX\braket[2]{\langle}{\rangle}{#1 \delimsize\vert #2}

\theoremstyle{definition}

\hypersetup{citecolor=blue}
\newcommand{\dif}{\mathrm{d}}
\newcommand{\Eqref}[1]{(\ref{#1})}
\newcommand{\half}{\frac{1}{2}}
\newcommand{\expo}[1]{\mathrm{e}^{#1}}
\newcommand{\brac}[1]{\left(#1 \right)}
\newcommand{\sbrac}[1]{\left[#1\right]}
\newcommand{\im}{\mathrm{i}}

\newcommand{\Dcal}{\mathcal{D}}
\newcommand{\Vcal}{\mathcal{V}}
\newcommand{\Acal}{\mathcal{A}}

\numberwithin{equation}{section}
 
\begin{document}

\title{Solenoid configurations and gravitational free energy of the AdS--Melvin spacetime}

\author[1]{Yen-Kheng Lim\footnote{Email: yenkheng.lim@gmail.com, yenkheng.lim@xmu.edu.my}}

\affil[1]{\normalsize{\textit{Department of Physics, Xiamen University Malaysia, 43900 Sepang, Malaysia}}}

\date{\normalsize{\today}}
\maketitle 

\renewcommand\Authands{ and }
\begin{abstract}
 In this paper we explore a solenoid configuration involving a magnetic universe solution embedded in an empty Anti-de Sitter (AdS) spacetime. This requires a non-trivial surface current at the interface between the two spacetimes which can be provided by a charged scalar field. When the interface is taken to the AdS boundary, we recover the full AdS--Melvin spacetime. The stability of the AdS--Melvin solution is also studied by computing the gravitational free energy from the Euclidean action.
\end{abstract}

\section{Introduction} \label{intro}

A \emph{magnetic universe} is a solution in Einstein--Maxwell theory which describes a configuration of magnetic field lines held together under its own gravity. One of the early considerations of this problem was by Wheeler \cite{Wheeler:1955} in the search of gravitational geons, and subsequent related solutions were found by Bonnor \cite{Bonnor:1954}. The form most relevant to the discussion of the present paper is by Melvin \cite{Melvin:1963qx}, and are commonly known as \emph{Melvin spacetimes}. In this paper we are interested in the counterpart to the Melvin spacetime that is asymptotic to Anti de-Sitter (AdS) spacetime, which we will refer to as the \emph{AdS-Melvin} spacetime. It is a solution to Einstein--Maxwell theory in the presence of a negative cosmological constant $\Lambda<0$. 

In the $\Lambda=0$ case, there were various methods to derive Melvin's original solution. One is to apply a Harrison transformation \cite{Harrison:1968} to a Minkowski seed. If a Schwarzschild black hole is taken as the seed, then the result of the Harrison transform is a black hole immersed in the Melvin universe \cite{Ernst:1975}. This procedure has also been generalised to higher dimensions by Ortaggio \cite{Ortaggio:2004kr}, and extended to include dilaton-type scalar fields by \cite{Dowker:1993bt,Galtsov:1998mhf,Yazadjiev:2005gs,Lim:2017dqw}. Alternatively, Havrdov\'{a} and Krtou\v{s} derives the Melvin solution by taking the charged C-metric (which describes a pair of charged accelerating black holes), and pushing the black hole far away while keeping the electromagnetic fields finite at the neighbourhood of the acceleration horizon \cite{Havrdova:2006gi}.

In the presence of a cosmological constant $\Lambda$, Astorino has derived the AdS--Melvin solution through a solution-generating method \cite{Astorino:2012zm}. The present author provided \cite{Lim:2018vbq} an analogue to Havrdov\'{a} and Krtou\v{s}'s procedure by taking the charged (A)dS C-metric \cite{Chen:2015vma} and pushing the black holes far away while keeping the electromagnetic fields finite near the acceleration horizon. 

In Ref.~\cite{Davidson:1999fa}, the authors considered a $\Lambda=0$ Melvin universe of finite radius is embedded in flat spacetime. This embedding requires a non-trivial stress-energy tensor at the surface separating the two distinct spacetimes. A suitable source was found to be a charged complex scalar field with an appropriate scalar potential. Physically this may be interpreted as a cylindrical current source that produces the Melvin fluxtube within it, and hence was dubbed the \emph{cosmic solenoid}. In this paper, we shall consider an AdS version of the solenoid by embedding the AdS--Melvin solution in a pure AdS background. By taking the solenoid radius to infinity, we recover the full AdS--Melvin solution. Henceforth, we shall use this terminology of \emph{full AdS--Melvin solution} to distinguish it from the \emph{AdS solenoid} of finite radius.

More recently, Kastor and Traschen \cite{Kastor:2020wsm} studied the geometrical and physical properties of the full AdS--Melvin solution. There are interesting differences between the AdS--Melvin solution and its $\Lambda=0$ counterpart. The AdS--Melvin solution is asymptotic to pure AdS spacetime, while the $\Lambda=0$ one is not asymptotically flat. Furthermore, for the AdS--Melvin solution, there exist a maximum magnetic flux $\Phi_{\mathrm{max}}$, under which there are two branches of AdS--Melvin solutions, which we denote by its magnetic field parameter $B_+$ and $B_-$. It was conjectured that one of these branches should be unstable. In this paper, we consider the thermodynamic approach by computing the gravitational free energy form the Euclidean action. We will see below that the free energy is proportional to $B^2$ and hence the $B_-$-branch has lower free energy and is thermodynamically favoured. We will also check the thermodynamic stability against the planar AdS black hole with a Ricci-flat horizon and obtain the parameters for a phase transition between the black hole and the AdS--Melvin solution.

The rest of this paper is organised as follows. In Sec.~\ref{sec_action} we present the action and equations of motion which governs our solutions. The AdS solenoid solution is constructed in Sec.~\ref{sec_solenoid}. Subsequently in Sec.~\ref{sec_thermodynamics} we consider the thermodynamic stability of the full AdS--Melvin solution. Conclusions and closing remarks are given in Sec.~\ref{sec_conclusion}.

\section{Action and equations of motion}\label{sec_action}

Consider a $D$-dimensional spacetime $M$ with a time-like hypersurface $\Sigma$, which partitions $M$ into two sides. We shall refer to $M^-$ as the `inner' side of $\Sigma$ and $M^+$ the `outer' side of $\Sigma$. We denote by $x^\mu$ the coordinates on $M^-$ with spacetime metric $\dif s_-^2=g_{\mu\nu}\dif x^\mu\dif x^\nu$, and $x_+^\mu$ the coordinates on $M^+$ with spacetime metric $\dif s_+^2=g^+_{\mu\nu}\dif x_+^\mu\dif x_+^\nu$. We define the surface $\Sigma$ as the boundary of $M^-$ with outward-pointing unit normal $n^\mu$, and the induced metric on $\Sigma$ is 
\begin{align}
 h_{\mu\nu}=g_{\mu\nu}-n_\mu n_\nu.
\end{align}
We shall denote by $y^a$ the intrinsic coordinates on $\Sigma$, such that the induced metric on $\Sigma$ is correspondingly
\begin{align}
 h_{ab}\dif y^a\dif y^b=h_{\mu\nu}e^\mu_a e^\nu_b,
\end{align}
where $e^\mu_a=\frac{\partial x^\mu}{\partial y^a}$.

We will consider an Einstein--Maxwell gravity with a cosmological constant $\Lambda$, where a gauge potential $A$ gives rise to a 2-form field $F=\dif A$. The action is given by
\begin{align}
 \begin{split}
 I&=I_{M^+}+I_{M^-}+I_\Sigma, \nonumber\\
 I_{M^\pm}&=\frac{1}{16\pi G}\int_{M^\pm}\dif^Dx\sqrt{-g}\;\brac{R-2\Lambda-{F}^2},\nonumber\\
  I_\Sigma&=\frac{1}{8\pi G}\oint_\Sigma\dif^{D-1}y\sqrt{-h}\brac{K^-+K^+}+I_{\psi}[h,\Acal,\psi],
 \end{split}
\end{align}
for some source field $\psi$, possibly coupled with the projected gauge potential $\Acal_a=A_\mu e^\mu_a$ on $\Sigma$. Here $R$ is the Ricci scalar, $F^2=F_{\mu\nu}F^{\mu\nu}$, and $K^\pm$ are the traces of the extrinsic curvatures $K^\pm_{\mu\nu}={h_\mu}^\lambda\nabla_\lambda n_\nu$ of $M^\pm$, respectively. Expressed in terms of intrinsic coordinates of $\Sigma$, the extrinsic curvatures are $K_{ab}^\pm=K_{\mu\nu}^\pm e^\mu_a e^\nu_b$. 

The Einstein--Maxwell equations in the bulk $M^\pm$ are
\begin{align}
 R_{\mu\nu}&=\frac{2\Lambda}{D-2}g_{\mu\nu}+2F_{\mu\lambda}{F_\nu}^\lambda-\frac{1}{D-2}F^2g_{\mu\nu},\\
 \nabla_\lambda F^{\lambda\nu}&=0,
\end{align}
For any tensorial quantity $T$ in the bulk, we use the notation $[T_{abc\cdots}]=T_{abc\cdots}^--T_{abc\cdots}^+$ to denote the jump of the quantity across $\Sigma$. The equations of motion on the surface $\Sigma$ are 
\begin{align}
 -\frac{1}{8\pi G}\brac{[K_{ab}]-[K]h_{ab}}&=\mathcal{T}_{ab}=\frac{2}{\sqrt{-h}}\frac{\delta I_\psi}{\delta h^{ab}},\\
 \frac{1}{4\pi G}[F_{\mu\nu}]n^\mu e^\nu_a&=\mathcal{J}_a=-\frac{1}{\sqrt{-h}}\frac{\delta I_\psi}{\delta \Acal^a}.
\end{align}
Where $\mathcal{T}_{ab}$\footnote{Note that we defined the surface stress tensor following the conventions of Brown and York \cite{Brown:1992br}, with the positive sign $\mathcal{T}_{ab}=+\frac{2}{\sqrt{-h}}\frac{\delta I_\psi}{\delta h^{ab}}$. The stress tensor defined in Ref.~\cite{Davidson:1999fa} comes with a negative sign, $S_{ab}=-\frac{2}{\sqrt{-h}}\frac{\delta I_\psi}{\delta h^{ab}}$, so one should use $S_{ab}=-\mathcal{T}_{ab}$ when comparing results in the literature.} and $\mathcal{J}_a$ are the surface stress tensor and surface current, respectively.

We choose the source to be a complex scalar field minimally coupled to $\Acal$ with the corresponding action 
\begin{align}
 I_\psi=-\frac{1}{8\pi G}\oint_\Sigma\dif^{D-1}y\sqrt{-h}\brac{\left|\Dcal\psi-\im e\Acal\psi\right|^2+\Vcal(|\psi|)},
\end{align}
where $\Dcal$ is the covariant derivative on $\Sigma$ compatible with $h_{ab}$, $e$ is the charge of the scalar field associated to its $U(1)$ symmetry, and $\left|\Dcal\psi-\im e\Acal\psi\right|^2=h^{ab}\brac{\Dcal_a\bar{\psi}+\im e\Acal_a\bar{\psi}}\brac{\Dcal_b\psi-\im e\Acal_b\psi}$. The scalar potential $\Vcal(|\psi|)$ is an appropriately chosen function of $|\psi|=\sqrt{\bar{\psi}\psi}$. The complex conjugate of $\psi$ is denoted by $\bar{\psi}$. 

For this action the surface stress tensor, surface current, and equation of motion for $\psi$ are respectively
\begin{align}
 \mathcal{T}_{ab}&=-\frac{1}{8\pi G}\sbrac{\brac{\Dcal_a\bar{\psi}+\im e\Acal_a\bar{\psi}}\brac{\Dcal_b\psi-\im e\Acal_b\psi}+(a\leftrightarrow b)-\brac{\left|\Dcal\psi-\im e\Acal\psi \right|^2+\Vcal}h_{ab}},\\
\mathcal{J}_a&=\frac{1}{8\pi G}\brac{\im e\bar{\psi}\Dcal_a\psi-\im e\psi\Dcal_a\bar{\psi}+2e^2\Acal_a|\psi|^2},\\
\frac{\psi}{2|\psi|}\frac{\dif\Vcal}{\dif|\psi|}&=\brac{\Dcal^c-\im e\Acal^c}\brac{\Dcal_c-\im e\Acal_c}\psi.
\end{align}

\section{AdS solenoid solution} \label{sec_solenoid}

To describe the AdS solenoid, we take the interior spacetime to be the AdS--Melvin magnetic universe,
\begin{subequations}
\begin{align}
 \dif s_-^2&=f(r)\dif\varphi^2+\frac{\dif r}{f(r)}+\frac{r^2}{\ell^2}\brac{-\dif t^2+\dif x_1^2+\ldots+\dif x^2_{D-3}},\label{AdSsol_metric}\\
    f(r)&=\frac{r^2}{\ell^2}-\frac{\mu}{r^{D-3}}-\frac{B^2}{r^{2(D-3)}}, \label{Mm_g}
\end{align}
\end{subequations}
where the gauge potential and its corresponding 2-form field is given by 
\begin{align}
 A=\sqrt{\frac{D-2}{2(D-3)}}\brac{\frac{B}{r_0^{D-3}}-\frac{B}{r^{D-3}}}\;\dif\varphi,\quad F^-=\sqrt{\half (D-2)(D-3)}\frac{B}{r^{D-2}}\;\dif r\wedge\dif\varphi. \label{Mm_A}
\end{align}
Here $B$ parametrises the strength of the magnetic field,  $\mu$ is regarded as the `soliton parameter', and $\ell$ is the AdS curvature scale related to the negative cosmological constant by  
\begin{align}
 \ell^2=-\frac{(D-1)(D-2)}{2\Lambda}. \label{AdS_l}
\end{align}
For the case $B=0$, the magnetic field vanishes and the solution reduces to that of the Horowitz--Myers soliton \cite{Horowitz:1998ha}.

The tip of the soliton is located at $r=r_0\geq0$, where $f(r_0)=0$. We shall call $r_0$  the \emph{soliton radius}. Therefore the potential shown in \Eqref{Mm_A} is chosen in the gauge where $A=0$ at the tip. The solution is symmetric under the simultaneous sign flips $B\rightarrow-B$ and $\varphi\rightarrow-\varphi$. Therefore without loss of generality we shall take $B\geq0$. It is then convenient to parametrise this family of solutions by $(r_0,B)$ and $\mu$ can be obtained from the parameters using $f(r_0)=0$ to write 
\begin{align}
 \mu=r_0^{D-3}\brac{\frac{r_0^2}{\ell^2}-\frac{B^2}{r_0^{2(D-3)}}}.
\end{align}

To ensure that the soliton caps off smoothly at $r_0$, the periodicity of the angular coordinate $\varphi$ shall be fixed to 
\begin{align}
 \Delta\varphi=\frac{2\pi}{\kappa},\quad\mbox{where}\quad \kappa=\half f'(r_0)=\frac{1}{2r_0}\sbrac{(D-1)\frac{r_0^2}{\ell^2}+(D-3)\frac{B^2}{r_0^{2(D-3)}}}. \label{Delta_phi}
\end{align}
The boundary of the spacetime will be $\Sigma$, located at $r=R$, with the outward-pointing unit normal
\begin{align}
 n=\sqrt{f(R)}\partial_r.
\end{align}
Therefore the coordinate range for the spacetime \Eqref{Mm_g} is taken to be $r_0\leq r\leq R$. Accordingly, the total magnetic flux contained in the region $r_0\leq r\leq R$ is
\begin{align}
 \Phi=\int_\gamma A=\Delta\varphi N, \label{Phi_B}
\end{align}
where the integration path is taken to be a circle of radius $r=R$, $\Delta\varphi$ is as given in \Eqref{Delta_phi}, and 
\begin{align}
 N=\sqrt{\frac{D-2}{2(D-3)}}\;B\brac{\frac{1}{r_0^{D-3}}-\frac{1}{R^{D-3}}}. \label{N_def}
\end{align}
Note that the flux depends quadratically on $B$. In Ref.~\cite{Kastor:2020wsm}, it was shown that there exists two $B$'s that give the same flux. In the present context of arbitrary $D\geq4$, we find it convenient to introduce the constants 
\begin{align}
 A_1=(D-1)\frac{r_0^2}{\ell^2},\quad A_2=(D-3)\frac{B^2}{r_0^{2(D-3)}},\quad \Delta=\sqrt{\frac{D-2}{2(D-3)}}\brac{\frac{1}{r_0^{D-3}}-\frac{1}{R^{D-3}}},
\end{align}
so that $\kappa=\frac{1}{2r_0}\brac{A_1+A_2B^2},\; N=B\Delta$, and therefore
\begin{align}
 \Phi=\frac{2\pi r_0 B\Delta}{A_1+A_2B^2}.
\end{align}
Solving for $B$ gives the two branches 
\begin{align}
 B_\pm&=\frac{2\pi r_0\Delta}{A_2\Phi}\brac{1\pm\sqrt{1-\frac{\Phi^2}{\Phi^2_{\mathrm{max}}}}}, 
\end{align}
where $\Phi_{\mathrm{max}}$ is the maximum flux for which the two branches join:
\begin{align}
 \Phi_{\mathrm{max}}=\frac{2\pi r_0\Delta}{\sqrt{A_1A_2}}=\frac{1}{D-3}\sqrt{\frac{2(D-2)}{D-1}}\;\pi\ell\brac{1-\frac{r_0^{D-3}}{R^{D-3}}}, \label{Phimax}
\end{align}
and the corresponding $B$ where this occurs is 
\begin{align}
 B_{\mathrm{max}}=\sqrt{\frac{A_1}{A_2}}=\sqrt{\frac{D-1}{D-3}}\frac{r_0^{D-3}}{\ell}.
\end{align}
In the limit $R\rightarrow\infty$, we recover maximum flux of the full AdS-Melvin spacetime. In particular, for $D=4$ and $R\rightarrow\infty$, we recover $\Phi_{\mathrm{max}}={2\pi\ell}/{\sqrt{3}}$, which is Eq.~(27) of \cite{Kastor:2020wsm}.

Turning to the exterior spacetime $M^+$, we shall take the pure AdS solution with the metric 
\begin{align}
 \dif s_+&=\frac{r^2}{\ell^2}C_\varphi\dif\varphi^2+\frac{\ell^2\dif r}{r^2}+\frac{r^2}{\ell^2}\brac{-C_t\dif t^2+C_1\dif x_1^2+\ldots+C_{D-3}\dif x_{D-3}^2}, \label{Pure_AdS}
\end{align}
where we have chosen our coordinates on $M^+$ to be 
\begin{align*}
 \varphi_+=\sqrt{C_\varphi}\varphi,\quad t_+=\sqrt{C_t}t,\quad x_{+i}=\sqrt{C_i}x_i,\quad i=1,\ldots,D-3.
\end{align*}
The constants $C_\varphi$, $C_t$, and $C_1,\ldots,C_{D-3}$ will be be chosen shortly such that the metric is continuous across $\Sigma$. The exterior spacetime will be taken to have zero magnetic field. Therefore the gauge potential is simply a constant. To ensure continuity of $A$ across $\Sigma$, we choose this constant to be
\begin{align}
 A=N\;\dif\varphi,\quad F^+=0,
\end{align}
where $N$ is as defined in \Eqref{N_def}.

The surface $\Sigma$ is the boundary of $M^+$ at $r=R$, this time with the inward-pointing normal 
\begin{align}
 n=\frac{r}{\ell}\partial_r,
\end{align}
so that the coordinate range for the exterior spacetime is $R\leq r<\infty$. Then, the continuity of the bulk metric across $\Sigma$ (at $r=R$ in the coordinates of both interior and exterior metrics) requires 
\begin{align}
 C_\varphi=\frac{\ell^2}{R^2}f(R),\quad C_t=C_1=\ldots=C_{D-3}=1,
\end{align}
and the jump of the trace of extrinsic curvature is 
\begin{align}
 [K]=\frac{D-2}{r}f(R)^{1/2}+\half f(R)^{-1/2}f'(R)-\frac{D-1}{\ell},
\end{align}
where primes denote derivatives with respect to $r$, so in our notation $f(R)$ and $f'(R)$ respectively denote the function $f(r)$ and its derivative evaluated at $r=R$. The surface stress tensor components are 
\begin{subequations} \label{T_ab}
\begin{align}
 \mathcal{T}_{ij}&=-\frac{1}{8\pi G}\brac{[K_{ij}]-[K]h_{ij}}=\frac{\sigma}{8\pi G}\brac{\frac{R^2}{\ell^2}\eta_{ij}},\\
 \mathcal{T}_{\varphi\varphi}&=-\frac{1}{8\pi G}\brac{[K_{\varphi\varphi}]-[K]h_{\varphi\varphi}}=\frac{\Omega}{8\pi G}f(R),
\end{align}
\end{subequations}
where  $\eta_{ij}\dif x^i\dif x^j=-\dif t^2+\dif x_1^2+\ldots+\dif x_{D-3}^2$ is the $(D-2)$-dimensional Minkowski metric and we have denoted
\begin{subequations}
\begin{align}
 \sigma&=\frac{D-3}{R}f(R)^{1/2}+\half f(R)^{-1/2}f'(R)-\frac{D-2}{\ell},\\
 \Omega&=\frac{D-2}{R}f(R)^{1/2}-\frac{D-2}{\ell}.
\end{align}
\end{subequations}
The surface current is
\begin{align}
 \frac{1}{4\pi G}[F_{\mu\nu}]n^\mu e_a^\nu=\frac{1}{4\pi G}\sqrt{\half(D-2)(D-3)}\frac{Bf(R)^{1/2}}{R^{D-2}}\;\delta^\varphi_a.
\end{align}

We take for $\psi$ the ansatz 
 \begin{align}
  \psi=\eta\expo{\im n\kappa \varphi},
 \end{align}
where $\eta$ is a constant and $n$ is an integer so that the complex phase is single-valued and periodic with angular periodicity $\Delta\varphi$ according to Eq.~\Eqref{Delta_phi}. For this ansatz, the surface stress tensor equations are 
\begin{align}
 \sigma=\Vcal+\frac{\eta^2}{f(R)}\brac{n\kappa-eN}^2,\quad\Omega=\Vcal-\frac{\eta^2}{f(R)}\brac{n\kappa-eN}^2.\label{eom_surfaceT}
\end{align}
The surface current equations become 
\begin{align}
 \sqrt{\half(D-2)(D-3)}\frac{B}{R^{D-2}}f(R)^{1/2}=\eta^2e(n\kappa-eN), \label{eom_surfaceJ}
\end{align}
and the Klein--Gordon equation for $\psi$ reduces to 
\begin{align}
 \frac{\dif\Vcal}{\dif|\psi|}=-\frac{2\eta}{f(R)}\brac{n\kappa-eN}^2.\label{eom_KG}
\end{align}
Eliminating $\eta$ from Eqs.~\Eqref{eom_surfaceT} and \Eqref{eom_surfaceJ} leads to the following equation:
\begin{align}
 2(D-2)(D-3)B^2\brac{n\kappa-eN}^2=e^2 R^{2(D-2)}(\sigma-\Omega)^2 f(R). \label{quadratic_e}
\end{align}
From \Eqref{eom_surfaceT} and \Eqref{eom_surfaceJ} one can also obtain the required value of $\eta$,
\begin{align}
 \eta^2=\frac{(D-2)(D-3)B^2}{e^2(\sigma-\Omega)R^{2(D-2)}}. \label{eta_eqn}
\end{align}
This equation tells us that $\sigma-\Omega$ must be positive. This can be checked by direct computation, as
\begin{align}
 \sigma-\Omega&=f(R)^{-1/2}\brac{\half f'(R)-\frac{1}{R}f(R)}\nonumber\\
   &=\frac{1}{R}f(R)^{-1/2}\sbrac{\frac{D-2}{2}\frac{\mu}{R^{D-3}}+(D-2)\frac{B^2}{R^{2(D-3)}}},
\end{align}
which is always positive away from the soliton tip.

\begin{figure}
 \begin{subfigure}[b]{0.49\textwidth}
    \centering
    \includegraphics[scale=0.9]{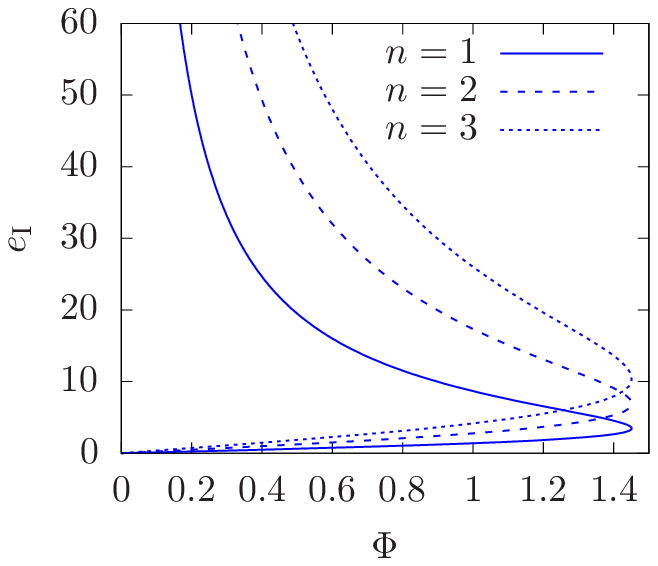}
    \caption{$R=5$.}
    \label{fig_ChooseR5}
  \end{subfigure}
  \begin{subfigure}[b]{0.49\textwidth}
    \centering
    \includegraphics[scale=0.9]{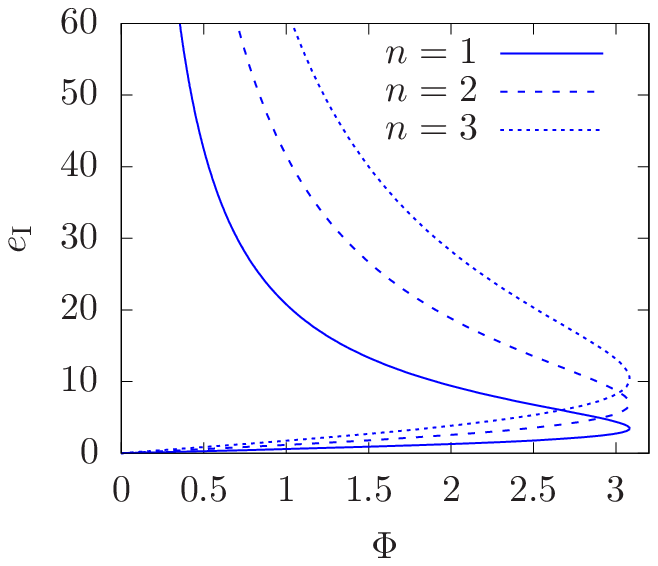}
    \caption{$R=20$.}
    \label{fig_ChooseR20}
  \end{subfigure}
  \caption{Plots of $\Phi$ vs $e_{\mathrm{I}}$ for $r_0=3$, in units where $\ell=1$.}
  \label{fig_ChooseR}
\end{figure}

Eqs.~\Eqref{quadratic_e} and \Eqref{eta_eqn}, determine the scalar charge $e$ and absolute value $\eta$ required to source an AdS-Melvin solenoid for a given radius $R$ and magnetic field parameter $B$. In particular, Eq.~\Eqref{quadratic_e} is a quadratic equation for $e$. The solution for general $D$ may appear cumbersome. However, for the case $D=4$, they are 
\begin{align}
 e_{\mathrm{I}}=\frac{2nB}{r_0^2},\quad\mbox{ or }\quad e_{\mathrm{II}}=\frac{2(r_0^4+B^2\ell^2)nBR}{r_0^2\brac{7B^2\ell^2 R-3Rr_0^4-8B^2\ell^2r_0}}.
\end{align}
For the second solution, we observe that for small $B$, 
\begin{align}
 e_{\mathrm{II}}\brac{n\kappa-e_{\mathrm{II}}N}\simeq-\frac{3n^2}{r_0\ell^2}B-\frac{2e_{\mathrm{II}}nr_0^2\ell^2\brac{r_0-R}}{Rr_0^5\ell^2}B^2+\mathcal{O}\brac{B^3}.
\end{align}
So this branch is continuously connected to a solution where $e_{\mathrm{II}}(n\kappa-e_{\mathrm{II}}N)$ is negative, which is in contradiction with Eq.~\Eqref{eom_surfaceJ}. Focussing our attention to the first solution, together with Eq.~\Eqref{Phi_B}, gives the charge required to produce an AdS solenoid of a given $B$, $R$, and $r_0$. Now, these latter three parameters also determine the total flux $\Phi$. Therefore we have a relationship between $\Phi$ and the required charge $e=e_{\mathrm{I}}$ of the scalar source. As a demonstrative example, Fig.~\ref{fig_ChooseR} shows the values of $(\Phi,e)$ for the case $\ell=1$, $r_0=3$, $R=5$, and $R=20$.  The curve starts at $e_{\mathrm{I}}=\Phi=0$ for $B=0$, and reaches a turning point for $\Phi=\Phi_{\mathrm{max}}$ and $B=B_{\mathrm{max}}$, given by Eq.~\Eqref{Phimax}. 

Finally, we check what kinds of scalar potential that would be able to support this solution. Suppose we take $\Vcal$ to have a quadratic form 
\begin{align}
 \Vcal(|\psi|)=\Vcal_0+\alpha|\psi|^2. \label{quadratic_V}
\end{align}
Then Eqs.~\Eqref{eom_surfaceJ} and \Eqref{eom_KG} lead to 
\begin{align}
 \Vcal_0=\sigma,\quad \alpha=-\frac{\brac{n\kappa-eN}^2}{f(R)}. \label{V_constants}
\end{align}
Since Eqs.~\Eqref{eom_surfaceT} and \Eqref{eom_KG} only determine the on-shell values of $\mathcal{V}$ and its first derivative, one is unable to claim that \Eqref{quadratic_V} is a unique potential that supports the AdS solenoid solution. However, restricting attention to $\mathcal{V}$ being a polynomial in $|\psi|$, the quadratic form \Eqref{quadratic_V} is perhaps the simplest choice.

To reiterate in the closing of this section, a surface complex scalar field of charge $e$ and magnitude $\eta$ is able to source the AdS solenoid of parameters $(B,r_0)$ and radius $R$ through Eqs.~\Eqref{quadratic_e} and \Eqref{eta_eqn}. A possible scalar potential that is able to support such a solution is \Eqref{quadratic_V}, where the numerical values of its constants should satisfy \Eqref{V_constants}.

\section{The full AdS-Melvin spacetime and Euclidean action} \label{sec_thermodynamics}

In this section we now take the surface $\Sigma$ to $R\rightarrow\infty$, so that $\Sigma=\partial M_-$ now becomes the AdS boundary containing the full AdS-Melvin solution in the bulk. In this context, Eq.~\Eqref{T_ab} now takes the interpretation of the boundary stress tensor \cite{Myers:1999psa} for the AdS-Melvin spacetime from which the contribution of pure AdS spacetime \Eqref{Pure_AdS} is subtracted. To leading order in $R$, the boundary stress tensor reads 
\begin{align}
 \hat{\mathcal{T}}_{ij}&=\frac{\mu\ell}{16\pi G R^{D-1}}\frac{R^2}{\ell^2}\eta_{ij},\quad  \hat{\mathcal{T}}_{\varphi\varphi}=-\frac{(D-2)\mu\ell}{16\pi G R^{D-1}}. \label{bdy_T}
\end{align}
In this context, the complex scalar $\psi$ merely serves as a physical model of a source for the stress tensor \Eqref{bdy_T}. When $R$ is taken to infinity at the end of the calculation, the surface containing $\psi$ is essentially pushed to infinity and no longer directly participates in the thermodynamic analysis below. 

Let $\xi$ be the time-like Killing vector of the spacetime. According to the Brown--York quasilocal stress tensor prescription \cite{Brown:1992br}, the energy of the spacetime is given by
\begin{align}
 M=\oint_B\dif^{D-2}x\sqrt{-h}\;\xi^\mu \xi^\nu\hat{\mathcal{T}}_{\mu\nu},
\end{align}
where $B$ is a hypersurface in $\partial M_-$ that is orthogonal to $\xi$. The result is 
\begin{align}
 M=\frac{\Delta\varphi\,\mathrm{vol}\brac{X}}{16\pi G \ell^{D-2}}\brac{\frac{B^2}{r_0^{D-3}}-\frac{r_0^{D-1}}{\ell^2}}, \label{AdSMel_M}
\end{align}
where $\Delta\varphi=\frac{2\pi}{\kappa}$ as defined in Eq.~\Eqref{Delta_phi} and 
\begin{align}
 \mathrm{vol}(X)=\int\dif x_1\cdots\dif x_{D-3},
\end{align}
which can be rendered finite by taking the coordinates $x_1,\ldots,x_{D-3}$ to have some finite periodicity.


To investigate the thermodynamics of the AdS-Melvin solution, we go to the Euclidean section by taking $t\rightarrow-\im\tau$, where $\tau$ is the Euclidean time with periodicity $\beta$. The Euclideanised AdS-Melvin solution is then
\begin{align}
 \dif s^2&=f\dif\varphi^2+\frac{\dif r}{f}+\frac{r^2}{\ell^2}\brac{\dif\tau^2+\dif x_1^2+\ldots+\dif x^2_{D-3}}, \label{Euclidean_AdSM}
\end{align}
where $f$, $\ell$, and $A$ are still as given as \Eqref{Mm_g}, \Eqref{Mm_A}, and \Eqref{AdS_l}.
This is a classical solution which extremises the Euclidean action
\begin{align}
 I_{\mathrm{E}}&=-\frac{1}{16\pi G}\int_{M_{\mathrm{E}}}\dif^D x\sqrt{g}\;\brac{R-2\Lambda-F^2}-\frac{1}{8\pi G}\oint_{\partial M_{\mathrm{E}}}\dif^{D-1}y\sqrt{h}\;K.
\end{align}
As is well known, computing the on-shell Euclidean action directly leads to a divergent result. Instead, we evaluate the action up to a finite boundary at $r=R$. Then we subtract the contribution of the pure AdS background where we can use \Eqref{Pure_AdS} but with $0\leq r\leq R$. The coordinate $\varphi$ was already scaled approprately using $C_\varphi$ so that the metrics match at $r=R$. Taking $R\rightarrow\infty$ towards the end, the result is
%
%
%
\begin{align}
 \hat{I}_{\mathrm{E}}&=\beta\frac{\Delta\varphi\mathrm{vol}(X)}{16\pi G\ell^{D-2}}\brac{\frac{B}{r_0^{D-3}}-\frac{r_0^{D-1}}{\ell^2}}\nonumber\\
                &=\beta M.
\end{align}
The gravitational free energy is simply $\mathcal{F}=\frac{\hat{I}_{\mathrm{E}}}{\beta}=M$, or explicitly in terms of parameters $(r_0, B)$,
\begin{align}
 \mathcal{F}&=\frac{\Delta\varphi\,\mathrm{vol}\brac{X}}{16\pi G \ell^{D-2}}\brac{\frac{B^2}{r_0^{D-3}}-\frac{r_0^{D-1}}{\ell^2}}.
\end{align}
As this is the free energy computed with pure AdS as the background, the pure AdS solution is one with zero free energy, $\mathcal{F}=0$. In other words, there is a critical value 
\begin{align}
 B_{\mathrm{crit}}=\frac{r_0^{D-2}}{\ell}, \label{B_crit_pure}
\end{align}
such that if $B<B_{\mathrm{crit}}$, the AdS-Melvin has free energy $\mathcal{F}<0$ and is thermodynamically favoured relative to pure AdS. On the other hand, for $B>B_{\mathrm{crit}}$, the free energy of AdS-Melvin is $\mathcal{F}>0$, and is thermodynamically unstable relative to pure AdS. For case $B=0$ we recover the fact that the Horowitz--Meyers soliton is thermodynamically stable against the pure AdS background.

As discussed in the previous section, and in \cite{Kastor:2020wsm}, recall that there are two branches of solutions of $B$ which gives rise to the same flux. The lower and upper branches correspond to $B_-<B_{\mathrm{max}}$ and $B_+>B_{\mathrm{max}}$, respectively, where $B_{\mathrm{max}}$ is given in Eq.~\Eqref{Phimax}. We find that 
\begin{align}
 \frac{B_{\mathrm{max}}}{B_{\mathrm{crit}}}=\sqrt{\frac{D-1}{D-3}}>1.
\end{align}
So $B_{\mathrm{max}}$ is always larger than $B_{\mathrm{crit}}$. In other words, the upper branch $B_+>B_{\mathrm{max}}$ is always in the thermodynamically unstable domain, and only the portion of the lower branch $B<B_{\mathrm{crit}}<B_{\mathrm{max}}$ is thermodynamically stable (relative to pure AdS).

Our discussions so far have been based on taking the pure AdS spacetimes as the background. On the other hand it is known that the planar AdS black hole is always stable against this background, and furthermore there exists a phase transition between the black hole and the Horowitz--Myers soliton \cite{Surya:2001vj}. Since AdS--Melvin solution introduces an additional parameter $B$ to the Horowitz--Meyers soliton, we should compare the Euclidean action against that of the AdS black hole.

The Euclideanised planar black hole solution is given by 
\begin{subequations}
\begin{align}
 \dif s^2&=\frac{r^2}{\ell^2}\tilde{C}_\varphi\dif\varphi^2+\frac{\dif r^2}{V(r)}+V(r)\tilde{C}_\tau\dif\tau^2+\frac{r^2}{\ell^2}\brac{\dif x_1^2+\ldots+\dif x^2_{D-3}},\\
  V(r)&=\frac{r^2}{\ell^2}-\frac{\nu}{r^{D-3}},
\end{align}
\end{subequations}
where $\nu$ is the mass parameter, and the constants $\tilde{C}_\varphi=\frac{\ell^2}{R^2}f(R)$ and $\tilde{C}_\tau=\frac{R^2}{\ell^2}\frac{1}{V(R)}$ are chosen to match with Eq.~\Eqref{Euclidean_AdSM} at the boundary $r=R$, which will be taken to infinity at the end. The black hole horizon is given by $r_+$, where $V(r_+)=0$. In the Euclidean section, the periodicity of the Euclidean time $\tau$ must be fixed to 
\begin{align}
 \beta=\frac{2\pi}{\kappa_{\mathrm{BH}}},\quad\mbox{ where }\quad\kappa_{\mathrm{BH}}=\frac{\sqrt{\tilde{C}_\tau}}{2}V'(r_+)
\end{align}
to avoid a conical singularity at $r=r_+$.

As before, we first evaluate the on-shell Euclidean actions of the AdS--Melvin $I_{\mathrm{E}}$ and the planar black hole $I_{\mathrm{E}}^{\mathrm{BH}}$ up to $R$. We then perform the subtraction $I_{\mathrm{E}}-I_{\mathrm{E}}^{\mathrm{BH}}$ and $R\rightarrow\infty$ is to be taken towards the end of the calculation. In this limit $\sqrt{\tilde{C}_\tau}\simeq 1$ and the black hole temperature is 
\begin{align}
 T=\beta^{-1}=\frac{(D-1)r_+}{4\pi\ell^2}.
\end{align}
The result for the background-subtracted action is
\begin{align}
 \hat{I}_{\mathrm{E}}=\beta\frac{\Delta\varphi\,\mathrm{vol}(X)}{16\pi G\ell^{D-2}}\brac{\frac{r_+^{D-1}}{\ell^2}+\frac{B^2}{r_0^{D-3}}-\frac{r_0^{D-1}}{\ell^2}},
\end{align}
and therefore the gravitational free energy is $\mathcal{F}=\frac{\hat{I}_{\mathrm{E}}}{\beta}$,
\begin{align}
 \mathcal{F}&=\frac{\Delta\varphi\,\mathrm{vol}(X)}{16\pi G\ell^{D-2}}\brac{\frac{r_+^{D-1}}{\ell^2}+\frac{B^2}{r_0^{D-3}}-\frac{r_0^{D-1}}{\ell^2}}.
\end{align}
As we are now comparing against the black-hole background, $\mathcal{F}<0$ means the AdS--Melvin is thermodynamically favoured, and $\mathcal{F}>0$ is where the black hole is favoured. The critical value occurs at 
\begin{align}
 B^2_{\mathrm{crit}}=\frac{r_0^{2(D-2)}-r_+^{D-1}r_0^{D-3}}{\ell^2}.
\end{align}
The presence of the parameter $r_+$ reduces the value of $B_{\mathrm{crit}}$ compared to \Eqref{B_crit_pure}. So when $B$ exceeds $B_{\mathrm{crit}}$, the AdS--Melvin spacetime becomes unstable relative to the planar black hole.

\section{Conclusion} \label{sec_conclusion}

In this paper we have explored a current source configuration that gives rise to an AdS--Melvin magnetic spacetime of finite radius embedded in pure AdS spacetime. The source takes the form of a complex chaged scalar field with an appropriately chosen potential. The equations of motion establishes relations between the scalar charge $e$, and magnitude $\eta$ with the solenoid radius $R$ and magnetic field parameter $B$. It was also shown that for a range of total flux, there exist two branches of solutions of distinct values of $B$. namely $B<B_{\mathrm{max}}$ and $B>B_{\mathrm{max}}$, where $B_{\mathrm{max}}$ is the point corresponding to maximum flux $\Phi_{\mathrm{max}}$.

When the solenoid radius is taken to infinity, we recover the full AdS--Melvin spacetime that is also asymptotically AdS. In this case, we have determined the gravitational free energy from its Euclidean action and compared it against the pure AdS and planar AdS black hole spacetimes. We find that, for a given flux $\Phi$, a portion of the branch with lower $B$, in the range $0\leq B<B_{\mathrm{crit}}$ is thermodynamically favoured, where $B_{\mathrm{crit}} <B_{\mathrm{max}}$. When the magnetic field parameter exceeds $B_{\mathrm{crit}}$, there exist a phase transiton to a planar AdS black hole.

\section*{Acknowledgements}

Y.-K.~L is supported by Xiamen University Malaysia Research Fund (Grant no. XMUMRF/ 2021-C8/IPHY/0001). 

\bibliographystyle{AdS-solenoid}

\bibliography{AdS-solenoid}

\end{document}